\documentclass[10pt, twoside, journal]{IEEEtran}

\usepackage{graphicx}
\usepackage{amsmath}
\usepackage{cuted}

\newcommand{\dyadic}[1]{{#1}
\setbox0=\hbox{$\scriptstyle\leftrightarrow$}
   \setbox2=\hbox{$#1$}
   \dimen0=.5\wd0 \advance\dimen0 by-.5\wd2
   \advance\dimen0 by-\wd0
   \kern\dimen0
{^{\raise.2ex\hbox{$\scriptstyle\leftrightarrow$}}}}

\newcommand{\dyadictall}[1]{{#1}
\setbox0=\hbox{$\scriptstyle\leftrightarrow$}
   \setbox2=\hbox{$#1$}
   \dimen0=.5\wd0 \advance\dimen0 by-.5\wd2
   \advance\dimen0 by-\wd0
   \kern\dimen0
{^{\raise.5ex\hbox{$\scriptstyle\leftrightarrow$}}}}

\DeclareFontFamily{U}{cmbsy}{}
\DeclareFontShape{U}{cmbsy}{m}{n}{ <5> <6> <7> <8> <9> gen * cmbsy
       <10> <10.95> <12> <14.4> <17.28> <20.74> <24.88> cmbsy10}{}
\DeclareMathAlphabet{\scb}{U}{cmbsy}{m}{n}

\begin{document}

\title{Average Transition Conditions for Electromagnetic Fields at a Metascreen of Vanishing Thickness}

\author{Edward~F.~Kuester,~\IEEEmembership{Life Fellow,~IEEE}, Enbo Liu and Nick J. Krull
        \thanks{Manuscript received \today.}\thanks{E. F. Kuester and N. J. Krull are with the Department of Electrical, Computer and Energy Engineering, University of Colorado, Boulder, CO 80309 USA (email: Edward.Kuester@colorado.edu). Enbo Liu is with the Department of Mechatronics, University of Electronic Science and Technology of China (UESTC), Chengdu, China.}}

\markboth{IEEE Transactions on Antennas and Propagation}{Transition Conditions for Metascreen}

\maketitle

\begin{abstract}
Using a dipole interaction model, we derive generalized sheet transition conditions (GSTCs) for electromagnetic fields at the surface of a metascreen consisting of an array of arbitrarily shaped apertures in a perfectly conducting screen of zero thickness. Use of the GSTCs permits modeling of structures containing perforated surfaces much more rapidly than is possible with full-wave numerical simulations. These conditions require that the period of the array be smaller than about a third of a wavelength in the surrounding media, and generalize many results previously presented in the literature. They are validated by comparison with results of finite-element modeling, and show excellent agreement when conditions of their derivation are satisfied.
\vspace{7mm}

\end{abstract}

\section{Introduction}

\IEEEPARstart{R}{ecent} years have seen a large body of research into the field of electromagnetic metamaterials, and in particular into metasurfaces, which are two-dimensional (surface) versions of engineered artificial magnetodielectric materials (see, e.~g., \cite{hk3}-\cite{barbuto}). Not only do metasurfaces offer the possibility of devices with lower loss than those based on bulk metamaterials, but they have found application in controlling reflection, transmission and polarization of waves; as filters, absorbers, sensors and beamformers; and performing the functions of focusing and field transformation.

In \cite{kmh}, equivalent boundary conditions were derived for a \emph{metafilm}---a particular kind of metasurface consisting of a planar array of isolated scatterers characterized by their polarizabilities. These equivalent boundary conditions are a special case of what are called \emph{generalized sheet transition conditions} or GSTCs \cite{senior}. Subsequent work has found many applications of this theory, summarized for example in \cite{hk3}.

\begin{figure}[ht]
 \centering
 \scalebox{0.8}{\includegraphics*{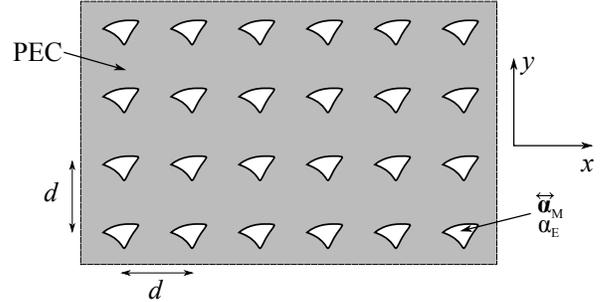}}
\caption{Metascreen consisting of a square array of identical apertures in a planar conducting screen.}
\label{f1}
\end{figure}

A complementary structure, to which we will give the name \emph{metascreen}, consists of an array of electrically small apertures in a planar conducting screen, as shown in Fig.~\ref{f1}.
Sakurai~\cite{sak} first obtained a form of GSTC for the case of a mesh of wires whose radius is small compared to the mesh openings. Later, Kontorovich and his colleagues \cite{kont2}-\cite{kont4} independently derived the same result and carried out extensive developments of that theory. Other approaches to the problem have considered the problem of plane-wave incidence onto the metascreen, representing it as an equivalent shunt reactance, calculated with a more or less approximated mode-matching technique \cite{mat1}-\cite{luuk} or finite-difference method \cite{kiley1}-\cite{kiley2}. For wire-mesh screens, a Bloch-Floquet-expansion-based numerical technique has been applied \cite{hill}-\cite{hill2}. Some of these results are restricted to the case of a normally incident plane wave, of identical media on both sides of the screen or to particular geometric forms of the metascreen. Of course, modern software modeling tools allow full-wave numerical solutions of the fields to be obtained, but these can require extensive computing time and resources, making the design of such surfaces inconvenient by comparison with the use of analytical models \cite{kiley1}-\cite{hill2}.

In the present paper, we will derive GSTCs that relate the average or macroscopic fields on a metascreen of fairly general form, restricted only by the requirements that the screen be a perfect conductor of zero thickness, and that the apertures are arranged in a square lattice whose period is small compared to a wavelength. In \cite{mshomog}, GSTCs for a metascreen of a fairly arbitrary geometry were derived by the method of multiple-scale homogenization, but computation of the coefficients appearing in the conditions obtained in this way requires the solution of several static boundary problems (which must be carried out numerically in general). Our approach here will be similar in some ways to that of \cite{lepp}-\cite{lepp2} for an analogous acoustical problem, and to that of \cite{bel}-\cite{bruna} for the problem of diffusion through a porous membrane. Latham \cite{lath} and Casey \cite{casey2} have partially carried out such an analysis for the electromagnetic problem, but have not obtained averaged boundary conditions for the screen. Delogne \cite{delogne} has outlined an approach that would lead to averaged boundary conditions, but presents explicit results only considering magnetic polarizabilities and neglecting the interaction of the apertures. The same comment applies to the work of Gordon \cite{rgordon}. A brief preliminary version of the present paper was presented in \cite{prelim}, after which the paper \cite{dimit} appeared. The latter has a certain degree of overlap with the present paper, but our work relies on a derivation independent of Babinet's principle and allows for different media on either side of the metascreen. The method of this paper is also extensible to a metascreen of nonzero thickness \cite{ekel}. These and other differences would seem to justify publication of this work.

\section{Derivation of the GSTCs}

\begin{figure}[ht]
 \centering
 \scalebox{0.9}{\includegraphics*{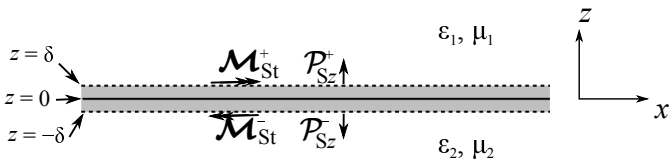}}
\caption{Side view of a metascreen showing equivalent surface polarization and magnetization densities.}
\label{f2}
\end{figure}

Consider the metascreen of Fig.~\ref{f1}. The screen is a perfect electric conductor (PEC) of zero thickness, in which a periodic square array of apertures is cut, whose lattice constant is $d$. Each aperture of the array in isolation has an electric polarizability $\alpha_E$ and a (dyadic) magnetic polarizability $\dyadic{\boldsymbol{\alpha}}_M$, in accordance with Bethe's small aperture theory. The definition of these polarizabilities is given in Appendix~\ref{apa}. When in the presence of a field, the effect of the apertures is to produce an additional field approximately equal to that produced by arrays of normal electric and tangential magnetic dipoles $\mathbf{p}_{\pm} = \mathbf{u}_z p_{z \pm}$ and $\mathbf{m}_{t \pm}$ located slightly above and below (at $z = \pm \delta$) a PEC screen with no holes (here $\mathbf{u}_v$ denotes a unit vector in the direction $v = x, y$ or $z$ of a cartesian coordinate system). These dipole arrays are in turn approximated by continuous distributions of surface polarization and magnetization densities:
\begin{equation}
 {\cal P}_{Sz}^{\pm} = N p_{z \pm} ; \qquad {\scb M}_{St}^{\pm} = N \mathbf{m}_{t \pm}
 \label{e1}
\end{equation}
where $N = 1/d^2$ is the density of apertures per unit area. The small distance $\delta$ will be allowed to approach zero later in our derivation. The resulting situation is shown in Fig.~\ref{f2}. The surface dipole densities on the bottom side of the metascreen are in opposite directions from those on the top side because of formulas (\ref{apa1})-(\ref{apa4}).

Now, between $z=0$ and $z=\delta$, and between $z=0$ and $z=-\delta$ (the shaded regions in Fig.~\ref{f2}), the tangential electric field will be zero because of the PEC at $z=0$. Because of the jump condition on tangential $\mathbf{E}$ across a sheet containing ${\cal P}_{Sz}^{\pm}$ and ${\scb M}_{St}^{\pm}$ (\cite{kmh,idemen}), we have
\begin{equation}
 \left. \mathbf{E} \right|_{z = \delta^+} \times \mathbf{u}_z = j \omega \mu_1 {\scb M}_{St}^+ - \nabla_t \left( \frac{{\cal P}_{Sz}^+}{\epsilon_1} \right) \times \mathbf{u}_z
\end{equation}
and
\begin{equation}
 \left. \mathbf{E} \right|_{z = -\delta^-} \times \mathbf{u}_z = - j \omega \mu_2 {\scb M}_{St}^- + \nabla_t \left( \frac{{\cal P}_{Sz}^-}{\epsilon_2} \right) \times \mathbf{u}_z
\end{equation}
Using (\ref{e1}) and (\ref{apa1})-(\ref{apa4}) from Appendix~\ref{apa}, then taking the limit as $\delta \rightarrow 0$, we obtain that tangential $\mathbf{E}$ is continuous at the metascreen:
\begin{equation}
 \left. \mathbf{E} \right|_{z = 0^+} \times \mathbf{u}_z = \left. \mathbf{E} \right|_{z = 0^-} \times \mathbf{u}_z \equiv \left. \mathbf{E} \right|_{z = 0} \times \mathbf{u}_z
 \label{econt}
\end{equation}
We emphasize that the electric field in equation (\ref{econt}) is not the actual field on the PEC between the two sheets of dipole densities, but the fields external to these dipole sheets ($z > \delta$ or $z < -\delta$), extrapolated to $z=0$. Tangential $\mathbf{E}$ also obeys the jump condition
\begin{eqnarray}
 \left. \mathbf{E} \right|_{z = 0} \times \mathbf{u}_z & = & j \omega \mu_{\rm av} N \dyadic{\boldsymbol{\alpha}}_M \cdot \left[ \mathbf{H}_t^{\rm sc} \right]_{z=0^-}^{0^+} \nonumber \\
 && + \frac{1}{\epsilon_{\rm av}} \nabla_t \left\{ N \alpha_E \left[ D_z^{\rm sc} \right]_{z=0^-}^{0^+} \right\} \times \mathbf{u}_z \label{ejump}
\end{eqnarray}
where
\begin{equation}
 \epsilon_{\rm av} = \frac{\epsilon_1 + \epsilon_2}{2} ; \qquad \mu_{\rm av} = \frac{2 \mu_1 \mu_2}{\mu_1 + \mu_2}
\end{equation}
and as detailed in Appendix~\ref{apa}, the \emph{short-circuit} fields $\mathbf{D}^{\rm sc}$, $\mathbf{B}^{\rm sc}$ are the fields acting on one of the apertures when that aperture is metalized. These fields are those produced by sources located on both sides of the screen, including those produced by all the other apertures.

\begin{figure}[ht]
 \centering
 \scalebox{0.8}{\includegraphics*{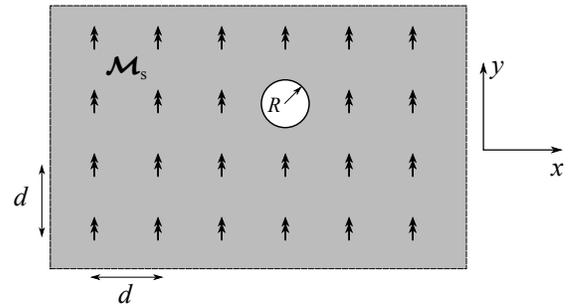}}
\caption{Field of magnetic dipole array approximated by that of a punctured sheet of magnetization density ${\scb M}_{S}$.}
\label{f2exc}
\end{figure}

We must now obtain suitable expressions for the short-circuit fields. Following the procedure used in \cite{kmh}, these fields at $z=0^{\pm}$ are equal to the macroscopic field $D_z$ or $\mathbf{H}_t$ at the metascreen, minus the fields of a disk of radius
\begin{equation}
 R = \frac{2\pi d}{\sum'_{m,n} (m^2 + n^2)^{3/2}} \simeq 0.6956 d
 \label{esquare}
\end{equation}
cut out of the surface polarization and magnetization sheets located at $z = \pm \delta$ respectively [the prime in (\ref{esquare}) denotes that the term with $m=n=0$ is to be omitted from the summation]. The situation is illustrated in Figure~\ref{f2exc} for the array of magnetic dipoles; a similar configuration holds for the electric dipoles.
In this procedure, the fields of all dipoles except those at the center of the exclusion circle have been approximated by the fields of the continuous distribution of surface magnetization density or surface polarization density with the disks removed. The fields of these disks are computed with the disks acting in the presence of the PEC screen at $z=0$ that has \emph{no apertures}. For example, the disk at $z=\delta$ acts above the PEC and produces a field that is the same as if the PEC were removed and its effect replaced with that of an image disk radiating together with the original disk in an unbounded region with material parameters $\mu_1$ and $\epsilon_1$. In other words, the field will be that produced by $2 {\scb M}_{St}^+$ and $2 {\cal P}_{Sz}^+$, calculated as in section III of \cite{kmh}.\footnote{
Note that eqn.~(48) of \cite{kmh} contains a typographical error; the last term on the right side should be
\[
 + j \frac{\omega {\cal P}_{sy}}{2} {\rm sgn}(z)
\]
but since in the present paper ${\cal P}_{sx} = {\cal P}_{sy} = 0$, this error has no effect here.}

From (47)-(49) of \cite{kmh}, we have
\begin{equation}
 \left. D_z^{\rm disk} \right|_{z = \delta^-} = 2 {\cal P}_{Sz}^+ F_1(R) ; \quad \left. \mathbf{H}_t^{\rm disk} \right|_{z = \delta^-} = 2 {\scb M}_{St}^+ G_1(R)
\end{equation}
where
\begin{eqnarray}
 F_1(R) & = & \frac{1}{2R} e^{-jk_1R} (1 + jk_1R) \nonumber \\
 & = & \frac{1}{2R} \left[ 1 + O(k_1^2 R^2) \right] \qquad (k_1R \ll 1)
\label{f1ap}
\end{eqnarray}
\begin{eqnarray}
 G_1(R) & = & - \frac{1}{4R} \left[ e^{-jk_1R} (1 - jk_1R) + 2jk_1R \right] \nonumber \\
 & = & - \frac{1}{4R} \left[ 1 + O(k_1^2 R^2) \right] \qquad (k_1R \ll 1)
\label{g1ap}
\end{eqnarray}
where $k_1 = \omega \sqrt{\mu_1 \epsilon_1}$. In a similar way,
\begin{equation}
 \left. D_z^{\rm disk} \right|_{z = -\delta^+} = 2 {\cal P}_{Sz}^- F_2(R) ; \ \ \left. \mathbf{H}_t^{\rm disk} \right|_{z = -\delta^+} = 2 {\scb M}_{St}^- G_2(R)
\end{equation}
where $F_2$ and $G_2$ are obtained from $F_1$ and $G_1$ by replacing the subscript 1 with 2. Therefore, letting $\delta^- \rightarrow 0^+$ and $-\delta^+ \rightarrow 0^-$, and assuming that $k_{1,2} R \ll 1$, we have finally
\begin{eqnarray}
 \left[ D_z^{\rm sc} \right]_{z=0^-}^{0^+} & = & \left[ D_z \right]_{z=0^-}^{0^+} - \left. D_z^{\rm disk} \right|_{z = 0^+} + \left. D_z^{\rm disk} \right|_{z = 0^-} \nonumber \\
 & = & \left[ D_z \right]_{z=0^-}^{0^+} - \frac{1}{R} \left( {\cal P}_{Sz}^+ - {\cal P}_{Sz}^- \right)
\end{eqnarray}
and
\begin{eqnarray}
 \left[ \mathbf{H}_t^{\rm sc} \right]_{z=0^-}^{0^+} & = & \left[ \mathbf{H}_t \right]_{z=0^-}^{0^+} - \left. \mathbf{H}_t^{\rm disk} \right|_{z = 0^+} + \left. \mathbf{H}_t^{\rm disk} \right|_{z = 0^-} \nonumber \\
 & = & \left[ \mathbf{H}_t \right]_{z=0^-}^{0^+} + \frac{1}{2R} \left( {\scb M}_{St}^+ - {\scb M}_{St}^- \right)
\end{eqnarray}

Now, from (\ref{e1}), (\ref{apa1}) and (\ref{apa3}) we have
\begin{equation}
 {\cal P}_{Sz}^+ - {\cal P}_{Sz}^- = - 2 N \alpha_E \left[ D_z^{\rm sc} \right]_{z=0^-}^{0^+}
\end{equation}
which gives
\begin{equation}
  \left[ D_z^{\rm sc} \right]_{z=0^-}^{0^+} = \left[ D_z \right]_{z=0^-}^{0^+} + \frac{2N}{R} \alpha_E \left[ D_z^{\rm sc} \right]_{z=0^-}^{0^+}
\end{equation}
or
\begin{equation}
  \left[ D_z^{\rm sc} \right]_{z=0^-}^{0^+} = \frac{1}{1 - \frac{2N}{R} \alpha_E} \left[ D_z \right]_{z=0^-}^{0^+}
  \label{e21}
\end{equation}
Similarly, from (\ref{e1}), (\ref{apa2}) and (\ref{apa4}) we have
\begin{equation}
 {\scb M}_{St}^+ - {\scb M}_{St}^- = 2 N \dyadic{\boldsymbol{\alpha}}_M \cdot \left[ \mathbf{H}_t^{\rm sc} \right]_{z=0^-}^{0^+}
\end{equation}
so that
\begin{equation}
  \left[ \mathbf{H}_t^{\rm sc} \right]_{z=0^-}^{0^+} = \left[ \mathbf{H}_t \right]_{z=0^-}^{0^+} +\frac{N}{R} \dyadic{\boldsymbol{\alpha}}_M \cdot \left[ \mathbf{H}_t^{\rm sc} \right]_{z=0^-}^{0^+}
\end{equation}
or
\begin{equation}
  \left[ \mathbf{H}_t^{\rm sc} \right]_{z=0^-}^{0^+} = \left[ \dyadictall{\mathbf{1}}_t - \frac{N}{R} \dyadic{\boldsymbol{\alpha}}_M \right]^{-1} \cdot \left[ \mathbf{H}_t \right]_{z=0^-}^{0^+}
  \label{e22}
\end{equation}
where $\dyadictall{\mathbf{1}}_t = \mathbf{u}_x \mathbf{u}_x + \mathbf{u}_y \mathbf{u}_y$ is the tangential identity dyadic. In the case where $\dyadic{\boldsymbol{\alpha}}_M$ is diagonal, (\ref{e22}) reduces to
\begin{equation}
  \left[ H_x^{\rm sc} \right]_{z=0^-}^{0^+} = \frac{1}{1 - \frac{N}{R} \alpha_M^{xx}} \left[ H_x \right]_{z=0^-}^{0^+}
  \end{equation}
and
\begin{equation}
  \left[ H_y^{\rm sc} \right]_{z=0^-}^{0^+} = \frac{1}{1 - \frac{N}{R} \alpha_M^{yy}} \left[ H_y \right]_{z=0^-}^{0^+}
  \end{equation}
  
Substituting (\ref{e21}) and (\ref{e22}) into (\ref{ejump}), we obtain as our GSTC that $\mathbf{E}_t$ is continuous at $z=0$, and:
\begin{eqnarray}
 \left. \mathbf{E} \right|_{z = 0} \times \mathbf{u}_z & = & j \omega \mu_{\rm av} \dyadic{\boldsymbol{\pi}}_{MS}^t \cdot \left[ \mathbf{H}_t \right]_{z=0^-}^{0^+} \nonumber \\
 && - \frac{1}{\epsilon_{\rm av}} \nabla_t \left\{ \pi_{ES}^{zz} \left[ D_z \right]_{z=0^-}^{0^+} \right\} \times \mathbf{u}_z \label{egstc}
\end{eqnarray}
where we have defined electric and magnetic \emph{surface porosities} of the metascreen as
\begin{equation}
 \pi_{ES}^{zz} = -\frac{N \alpha_E}{1 - 2 \frac{N}{R} \alpha_E}
 \label{epies}
\end{equation}
and
\begin{equation}
 \dyadic{\boldsymbol{\pi}}_{MS}^t = N \dyadic{\boldsymbol{\alpha}}_M \cdot \left[ \dyadictall{\mathbf{1}}_t - \frac{N}{R} \dyadic{\boldsymbol{\alpha}}_M \right]^{-1} \label{epims}
\end{equation}
If the magnetic polarizability dyadic is diagonal, we can simplify (\ref{epims}) to
\begin{equation}
 \dyadic{\boldsymbol{\pi}}_{MS}^t = \mathbf{u}_x \mathbf{u}_x \frac{N \alpha_M^{xx}}{1 - \frac{N}{R} \alpha_M^{xx}} + \mathbf{u}_y \mathbf{u}_y \frac{N \alpha_M^{yy}}{1 - \frac{N}{R} \alpha_M^{yy}} \label{epimsdiag}
\end{equation}
The minus sign in (\ref{epies}) is chosen to achieve a certain duality in the form of (\ref{egstc}) when compared to the GSTCs of a metafilm. A consequence of this sign choice is that $\pi_{ES}^{zz}$ will be negative. The fact that $\dyadic{\boldsymbol{\pi}}_{MS}^t$ has only tangential components, while $\pi_{ES}^{zz}$ has only normal components is a consequence of the fact that tangential electric dipoles and normal magnetic dipoles placed on a perfectly conducting surface produce no external fields.

Equation (\ref{egstc}) has the same form as eqn.~(3) of \cite{hk3}, and conforms with that of \cite{hkdrs} for a wire grating. It can be shown that (\ref{econt}) and (\ref{egstc}) are also special cases of the GSTCs derived in \cite{mshomog} if the screen thickness is set equal to zero. It must be emphasized that this form of the GSTCs applicable to a metascreen differs in a fundamental way from those that have been derived for a metafilm in \cite{kmh}, and used extensively since then (see, e.~g., \cite{ach}-\cite{vaha}). A metafilm consists of an array of unconnected scatterers that can be modeled as polarization and magnetization sheets that cause discontinuities in both tangential $\mathbf{E}$ and tangential $\mathbf{H}$ that are proportional to those surface polarizations. Equation (\ref{egstc}) on the other hand expresses the tangential electric field itself in terms of the jumps of the macroscopic fields (along with the derivative of one of them), and is not expressible in the form of a GSTC for a metafilm.

Thus, even though our derivation of (\ref{egstc}) has been based on the dipole-interaction model (analogous to the Clausius-Mossotti-Lorentz-Lorenz model of dielectric permittivity), we see that (\ref{egstc}) will hold even without that restriction, but that expressions (\ref{epies})-(\ref{epimsdiag}) for the surface porosities will no longer be true in general. Indeed, (\ref{egstc}) has the same form as the boundary condition obtained by Sakurai \cite{sak}, though his results apply to the case of a mesh of thin wires for which (\ref{epies})-(\ref{epimsdiag}) cannot be expected to hold. We obtain expressions for $\pi_{ES}^{zz}$ and $\dyadic{\boldsymbol{\pi}}_{MS}^t$ for some specific metascreen geometries in Appendix~\ref{apb}, valid in both the dipole-interaction approximation, as well as in the limit of a thin-wire mesh.

It should be emphasized that we should not expect the GSTC (\ref{egstc}) to be accurate if the lattice constant $d$ is too large. Not only have we approximated the interaction between apertures by invoking the quasistatic approximations (\ref{f1ap})-(\ref{g1ap}), but the very form of the GSTC itself precludes the presence of propagating higher-order Bloch-Floquet modes on the lattice, which means that we should restrict the lattice constant to be less than half a wavelength.

We may convert our GSTC into a somewhat different form by defining the surface current density
\begin{equation}
 \mathbf{J}_S = \mathbf{u}_z \times \left[ \mathbf{H}_t \right]_{z=0^-}^{0^+}
\end{equation}
and using the result
\begin{equation}
 D_z = - \frac{1}{j\omega} \nabla_t \cdot \left( \mathbf{u}_z \times \mathbf{H}_t \right)
\end{equation}
that follows from Amp\`{e}re's law, so that (\ref{egstc}) can be expressed as
\begin{equation}
 \left. \mathbf{E}_t \right|_{z = 0} = j \dyadictall{\mathbf{X}}_{ms} \cdot \mathbf{J}_S - j \nabla_t \left[ \frac{\pi_{ES}^{zz}}{\omega \epsilon_{\rm av}} \nabla_t \cdot \mathbf{J}_S \right] \label{gstckontform}
\end{equation}
where
\begin{eqnarray}
 \lefteqn{\dyadictall{\mathbf{X}}_{ms} = \omega \mu_{\rm av} \left( \mathbf{u}_x \mathbf{u}_x \pi_{MS}^{yy} \right.} \\
 && \left. \mbox{} - \mathbf{u}_x \mathbf{u}_y \pi_{MS}^{yx} - \mathbf{u}_y \mathbf{u}_x \pi_{MS}^{xy} + \mathbf{u}_y \mathbf{u}_y \pi_{MS}^{xx} \right) \nonumber
\end{eqnarray}
is the dyadic surface reactance of the metascreen. Equation (\ref{gstckontform}) has the form of the boundary condition obtained by Kontorovich and his colleagues \cite{kont2}-\cite{kont4} for a thin-wire mesh, again with different expressions for the surface porosities than (\ref{epies})-(\ref{epimsdiag}).

A boundary condition on the normal component of $\mathbf{B}$ can be obtained by taking the tangential divergence of $\mathbf{E} \times \mathbf{u}_z$ and using a vector identity together with Faraday's law:
\begin{equation}
 \nabla_t \cdot \left( \mathbf{E} \times \mathbf{u}_z \right) = \mathbf{u}_z \cdot \nabla_t \times \mathbf{E} = -j\omega B_z
 \label{efar}
\end{equation}
From this, we see that $B_z$ must be continuous at the metascreen, while applying the tangential divergence to both sides of (\ref{egstc}) and employing a further vector identity gives:
\begin{equation}
 \mathbf{u}_z \cdot \nabla_t \times \left. \mathbf{E} \right|_{z = 0} = j \omega \mu_{\rm av} \nabla_t \cdot \left\{ \dyadic{\boldsymbol{\pi}}_{MS}^t \cdot \left[ \mathbf{H}_t \right]_{z=0^-}^{0^+} \right\} \label{e40}
\end{equation}
Applying (\ref{efar}) to (\ref{e40}) and dividing the result by $j\omega$ if the frequency is not zero, we obtain
\begin{equation}
 \left. B_z \right|_{z = 0} = - \mu_{\rm av} \nabla_t \cdot \left\{ \dyadic{\boldsymbol{\pi}}_{MS}^t \cdot \left[ \mathbf{H}_t \right]_{z=0^-}^{0^+} \right\} \label{egstcmstat}
\end{equation}
which has the same form as eqn.~(12) of \cite{casey}.

To summarize the main results of this paper, the macroscopic electromagnetic field at a perfectly conducting metascreen of zero thickness must obey the following conditions:
\begin{itemize}
 \item{1.} The tangential components of $\mathbf{E}$ and the normal component of $\mathbf{B}$ are continuous across the metascreen.
 \item{2.} The field must obey (\ref{egstc}) and (\ref{egstcmstat}) at the metascreen.
\end{itemize}

\section{The Static Limit}

In the electrostatic limit, we let $\omega \rightarrow 0$ in (\ref{egstc}) to obtain
\begin{equation}
 \left. \mathbf{E} \right|_{z = 0} \times \mathbf{u}_z = - \frac{1}{\epsilon_{\rm av}} \nabla_t \left\{ \pi_{ES}^{zz} \left[ D_z \right]_{z=0^-}^{0^+} \right\} \times \mathbf{u}_z \label{e30}
\end{equation}
A static electric field can be expressed in terms of a scalar potential, $\mathbf{E} = - \nabla \Phi$, and the condition that tangential $\mathbf{E}$ must be continuous at $z=0$ means that $\Phi(x,y,z=0^+) - \Phi(x,y,z=0^-)$ must be a constant, which we can choose to be zero with no loss of generality so that $\Phi$ is continuous at $z=0$, and equal to a function $\Phi_0(x,y)$ there. Provided that $\Phi_0$ and $D_z$ are not independent of $x$ and $y$, we have the electrostatic form of the GSTC:
\begin{equation}
 \Phi_0 = \frac{\pi_{ES}^{zz}}{\epsilon_{\rm av}} \left[ D_z \right]_{z=0^-}^{0^+}
 \label{egstcestat}
\end{equation}
where an arbitrary additive constant in the potential has been chosen to give the indicated value at the metascreen. If $\Phi_0$ and $D_z$ are constant, equation (\ref{e30}) simply states that $0=0$ and nothing further can be deduced from it, but if we take the limit of (\ref{egstcestat}) as $\Phi_0$ approaches a constant function, we can regard it as applying to this case as well. Equation~(\ref{egstcestat}) has the same form as equation~(25) of \cite{casey}.

\begin{figure}[ht]
 \centering
 \scalebox{0.91}{\includegraphics*{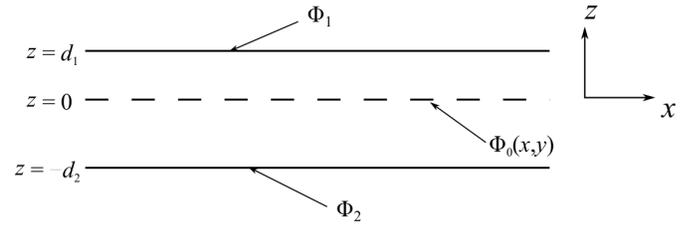}}
\caption{Metascreen between two conducting plates.}
\label{fstat}
\end{figure}

Suppose now that we place the metascreen between two conducting plates at $z=d_1$ and $z= -d_2$ as shown in Fig.~\ref{fstat}.
In $0<z<d_1$, the electric field is
\begin{equation}
 E_z = E_1 = \frac{1}{d_1} \left( \Phi_0 - \Phi_1 \right)
 \label{e31}
\end{equation}
where $\Phi_1$ is the potential at the upper plate, while in $-d_2 < z < 0$,
\begin{equation}
 E_z = E_2 = \frac{1}{d_2} \left( \Phi_2 - \Phi_0 \right)
 \label{e32}
\end{equation}
Applying the boundary condition (\ref{egstcestat}) together with (\ref{e31})-(\ref{e32}) and solving for the potential at the metascreen gives
\begin{equation}
 \Phi_0 = \frac{-\tilde{D}_1 + \tilde{D}_2}{\frac{\epsilon_1}{d_1} + \frac{\epsilon_2}{d_2} - \frac{\epsilon_{\rm av}}{\pi_{ES}^{zz}}}
\end{equation}
where
\begin{equation}
 \tilde{D}_1 = - \epsilon_1 \frac{\Phi_1}{d_1} ; \qquad \tilde{D}_2 = \epsilon_2 \frac{\Phi_2}{d_2}
\end{equation}
are the values of $D_z$ that would exist in $z>0$ and $z<0$ if the metascreen holes were metalized ($\left. \Phi \right|_{z=0} = 0$). If we compare this result for the case when $\epsilon_1 = \epsilon_2$ to eqn.~(16) of Grosser and Schulz \cite{grosser}, we see that in their notation $\delta \varphi \rightarrow \left. \Phi \right|_{z=0}$ and $b \rightarrow -2 \pi_{ES}^{zz}$. In particular, Table 1 of \cite{grosser} can now supply us with values of $\pi_{ES}^{zz}$ for a variety of aperture shapes.

The magnetostatic limit of our GSTC is (\ref{egstcmstat}) along with the continuity of $B_z$ at $z=0$, since $\omega$ does not appear explicitly in either condition.

\section{Equivalent Circuit}

Some researchers prefer to describe metasurfaces using surface impedances (see, e.~g., \cite{tret}). A simple impedance equivalent circuit for the metascreen of zero thickness can be obtained in certain special cases. Suppose that the field has no variation in the $y$-direction ($\partial/\partial y \equiv 0$) and that all fields vary with $x$ as $e^{-j k_x x}$. Then, as is well known, the field can be written as the superposition of a TE part (consisting of the field components $E_y$, $H_x$ and $H_z$ only) and a TM part (consisting of the field components $H_y$, $E_x$ and $E_z$ only). If we suppose in addition that the magnetic porosity dyadic is diagonal: $\dyadic{\boldsymbol{\pi}}_{MS}^t = \mathbf{u}_x \mathbf{u}_x \pi_{MS}^{xx} + \mathbf{u}_y \mathbf{u}_y \pi_{MS}^{yy}$, the metascreen will produce no conversion between these two polarizations, and they may be modeled independently of each other.

For the TE field, let $E_y \rightarrow V$ and $H_x \rightarrow -I$. Since $E_y$ is continuous at $z=0$, the GSTC (\ref{egstc}) can be interpreted as a shunt reactance $X_{\rm TE}$ at $z=0$, where
\begin{equation}
 X_{\rm TE} = \omega \mu_{\rm av} \pi_{MS}^{xx}
 \label{xte}
\end{equation}
Likewise, for the TM field let $E_x \rightarrow V$ and $H_y \rightarrow I$. By Amp\`{e}re's law and the assumptions above about the $x$- and $y$-dependences of the field, we have $E_z = -(k_x/\omega \epsilon) H_y$. Thus the GSTC (\ref{egstc}) in this case is equivalent to a shunt reactance at $z=0$ of
\begin{equation}
 X_{\rm TM} = \omega \mu_{\rm av} \pi_{MS}^{yy} + \frac{k_x^2 \pi_{ES}^{zz}}{\omega \epsilon_{\rm av}}
 \label{xtm}
\end{equation}
We should note that $jX_{\rm TE}$ and $jX_{\rm TM}$ are often called the transfer impedance when describing braided shields of, for instance, coaxial cables (see \cite{ikrath}-\cite{del1} and \cite{delogne}).

\section{Plane Wave Reflection and Transmission}

In this section, we will apply the GSTCs obtained above to the determination of the reflection and transmission coefficients of a plane wave incident on a metascreen. As in the previous section, the magnetic porosity dyadic will be assumed to be diagonal.
\begin{figure}[ht]
 \centering
 \scalebox{0.5}{\includegraphics*{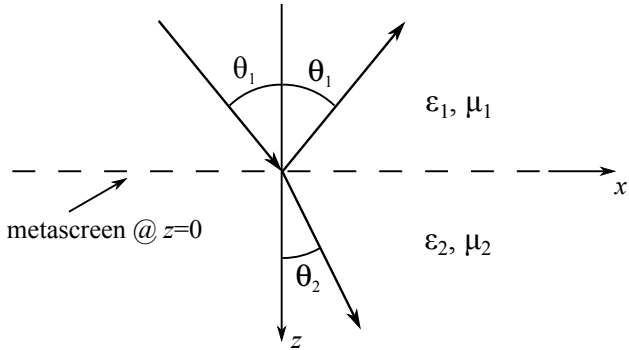}}
\caption{Plane wave incident at a metascreen.}
\label{f4}
\end{figure}

If a TE (perpendicular) polarized plane wave is incident at an angle $\theta_1$ to the $z$-axis as shown in Figure~\ref{f4}, the electric field $\mathbf{E} = \mathbf{u}_y E_y$ is given by
\begin{equation}
\begin{array}{l}
 E_y = e^{-jk_1 x \sin \theta_1} \left[ e^{-jk_1 z \cos \theta_1} + \Gamma_{\rm TE} e^{jk_1 z \cos \theta_1} \right] \ \mbox{\rm ($z<0$)} \\
 \quad \ \, = e^{-jk_2 x \sin \theta_2} T_{\rm TE} e^{-jk_2 z \cos \theta_2} \ \mbox{\rm ($z>0$)}
 \end{array}
\end{equation}
where $k_{1,2} = \omega \sqrt{\mu_{1,2} \epsilon_{1,2}}$, $\Gamma_{\rm TE}$ is the reflection coefficient, $T_{\rm TE}$ is the transmission coefficient, and the transmitted angle $\theta_2$ is related to the incident angle by Snell's law:
\begin{equation}
 k_1 \sin \theta_1 = k_2 \sin \theta_2 = k_x
 \label{snell}
\end{equation}
The magnetic field is obtained from Faraday's law $\nabla \times \mathbf{E} = - j\omega \mu \mathbf{H}$. Enforcing continuity of $E_y$ and the GSTC (\ref{egstc}) at $z=0$ in the usual way leads to the following formulas for the reflection and transmission coefficients:
\begin{equation}
 \Gamma_{\rm TE} = - \frac{1 - jX_{\rm TE} \left( \frac{\cos \theta_1}{\zeta_1} - \frac{\cos \theta_2}{\zeta_2} \right)}{1 + jX_{\rm TE} \left( \frac{\cos \theta_1}{\zeta_1} + \frac{\cos \theta_2}{\zeta_2} \right)}
 \label{gammaTE}
\end{equation}
and
\begin{equation}
 T_{\rm TE} = \frac{2jX_{\rm TE} \frac{\cos \theta_1}{\zeta_1}}{1 + jX_{\rm TE} \left( \frac{\cos \theta_1}{\zeta_1} + \frac{\cos \theta_2}{\zeta_2} \right)}
 \label{TTE}
\end{equation}
where $X_{\rm TE}$ is given by (\ref{xte}) and $\zeta_{1,2} = \sqrt{\mu_{1,2}/\epsilon_{1,2}}$ are the wave impedances of the upper and lower half-spaces. This result could of course have also been obtained by using the equivalent shunt reactance (\ref{xte}) connected across the junction of two transmission lines with characteristic impedances $\zeta_1/\cos \theta_1$ and $\zeta_2/\cos \theta_2$. It will be readily observed that when $\pi_{MS}^{xx} \rightarrow 0$, we obtain $\Gamma_{\rm TE} = -1$ and $T_{\rm TE} = 0$ consistent with an unperforated PEC screen, while for $\pi_{MS}^{xx} \rightarrow \infty$ (meaning that the metalization is removed), we retrieve the Fresnel coefficients for perpendicular polarization.

The reflection and transmission coefficients for a TM (parallel) polarized incident wave are derived in a similar manner, with somewhat more complicated expressions arising due to the presence of a normal component of the electric field. We obtain:
\begin{equation}
 \Gamma_{\rm TM} = - \frac{1 - jX_{\rm TM} \left( \frac{1}{\zeta_1 \cos \theta_1} - \frac{1}{\zeta_2 \cos \theta_2} \right)}{1 + jX_{\rm TM} \left( \frac{1}{\zeta_1 \cos \theta_1} + \frac{1}{\zeta_2 \cos \theta_2} \right)}
\end{equation}
and
\begin{equation}
 T_{\rm TM} = \frac{\frac{2jX_{\rm TM}}{\zeta_1 \cos \theta_2}}{1 + jX_{\rm TM} \left( \frac{1}{\zeta_1 \cos \theta_1} + \frac{1}{\zeta_2 \cos \theta_2} \right)}
 \label{TTM}
\end{equation}
where $X_{\rm TM}$ given by (\ref{xtm}) with $k_x$ is given by (\ref{snell}). Again, this result could have been obtained by using the equivalent shunt reactance (\ref{xtm}) connected across the junction of two transmission lines, this time with characteristic impedances $\zeta_1 \cos \theta_1$ and $\zeta_2 \cos \theta_2$. We obtain the appropriate limits if we allow the porosities to approach zero or infinity, with the exception of the special angle of incidence
\begin{equation}
\theta_1 = \arcsin \left(\sqrt{-\frac{\pi_{MS}^{yy}}{\pi_{ES}^{zz}} \frac{\mu_{\rm av} \epsilon_{\rm av}}{\mu_1 \epsilon_1}}  \right)
\end{equation}
for which case we have $\Gamma_{\rm TM} = -1$.

Equations (\ref{gammaTE})-(\ref{TTM}) agree with various results previously given in the literature if certain approximated forms of $X_{\rm TE, TM}$ are used---for normal incidence at a screen in a nonmagnetic material interface \cite{compt,whit}; and for oblique incidence at a wire mesh in free space \cite{kont2,ast2}. The expression given in \cite{rgordon} for the normal-incidence transmission coefficient differs from our result by a factor of 2 in the magnetic surface porosity; we believe our formula to be correct, as validated by the numerical results in the next section. The case of oblique incidence at a wire mesh at a height $h$ above a material half-space has been treated in \cite{spirina,ast4}, \cite[sect.~3.3]{kont4}, but these results cannot be reduced to our situation by letting $h \rightarrow 0$, because the near-field interaction of the mesh with the half-space is not accounted for in these works.

\section{Some Numerical Results}

\begin{figure}[ht!]
 \centering
 \scalebox{0.9}{\includegraphics*{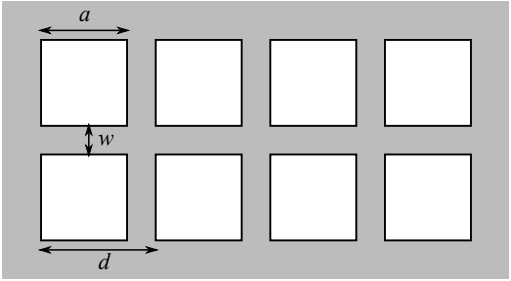}}
\caption{Metascreen with square apertures.}
\label{fsquare}
\end{figure}

Aside from checking limiting cases and special cases investigated by other authors, the accuracy of (\ref{gammaTE})-(\ref{TTM}) using (\ref{xte}) and (\ref{xtm}) can be assessed by comparing them to the results of full-wave numerical simulation. Our model will be a metascreen with square apertures as shown in Fig.~\ref{fsquare}.
In Appendix~\ref{apb}, we have obtained the uniformly valid (for any ratio of $a/d$) expressions (\ref{apc2}) and (\ref{apc1}) for the electric and magnetic surface porosities of this structure. These will be used in (\ref{xte})-(\ref{xtm}) and (\ref{gammaTE})-(\ref{TTM}) to obtain GSTC-based results for reflection and transmission.

\begin{figure}[ht!]
 \centering
 \scalebox{1.1}{\includegraphics*{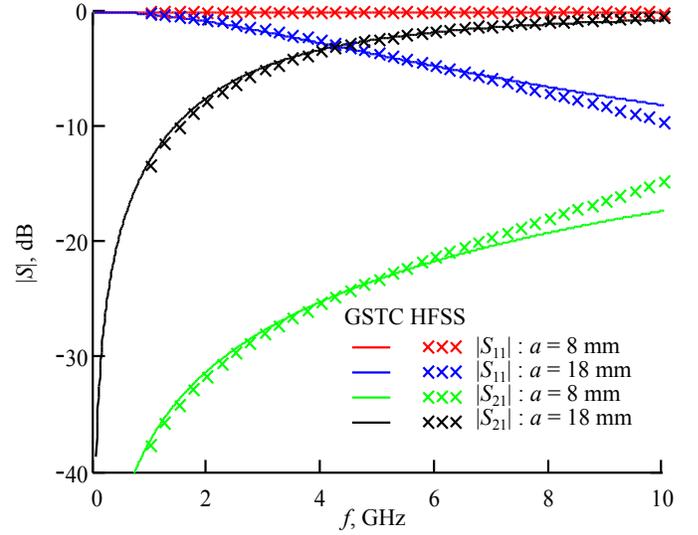}} \\
 (a) \\
 \vspace*{0.1in}
 \scalebox{1.1}{\includegraphics*{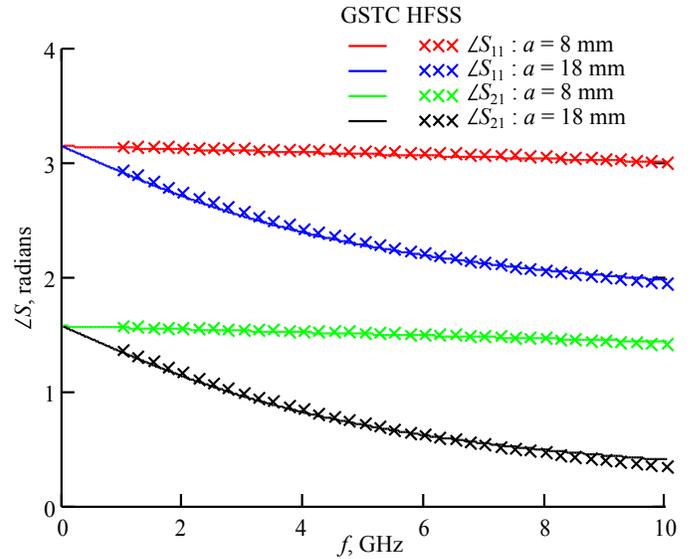}} \\
 (b)
\caption{Normal incidence reflection and transmission coefficients from the metascreen of Fig.~\ref{fsquare} with $d=20$ mm, $\epsilon_1 = \epsilon_2 = \epsilon_0$ and $\mu_1 = \mu_2 = \mu_0$: (a) magnitude, (b) phase.}
\label{fnorm}
\end{figure}

In Fig.~\ref{fnorm}, we present a comparison of the predictions of the present paper to finite-element simulations of normal-incidence plane wave reflection and transmission coefficients for the case when the medium on each side of the screen is free space. In this case, $\Gamma_{\rm TE} = \Gamma_{\rm TM} = S_{11}$ and $T_{\rm TE} = T_{\rm TM} = S_{21}$. The finite-element simulations were carried out in ANSYS HFSS, using pairs of PEC and PMC side walls at the boundaries of the period cell to force the normally incident plane waves at the metascreen.
We have chosen the lattice constant to be $d = 20$ mm, which is equal to a half wavelength at $f_{\lambda/2} = 7.5$ GHz. We can see that the magnitudes and phases of the transmission coefficient show good agreement between the GSTC prediction and numerical results up to $f_{\lambda/2}$, and the phases of the reflection and transmission coefficients are quite accurate well above that frequency. Of course, at normal incidence, the higher-order Floquet mode that begins to propagate above $f_{\lambda/2}$ is not excited due to symmetry, and it is to be expected that good agreement will be obtained until we near the next Floquet-mode threshold at $2f_{\lambda/2} = 15$ GHz. The finite-element solution was carried out with as fine a mesh as reasonably possible on a PC with 8 GB of memory; this typically resulted in more than 300,000 first-order elements, and small (less than 0.05\%) variations in computed $S$-parameters as the mesh was refined to its final size, so high accuracy can be attributed to these results, although when $|S_{11}|$ or $|S_{21}|$ becomes numerically small, the small discretization errors in the finite-element simulation seem to magnify the discrepancies observed on the dB scale in Figure~\ref{fnorm}(a), especially as the frequency begins to approach 15 GHz.

\begin{figure}[ht!]
 \centering
 \scalebox{1.2}{\includegraphics*{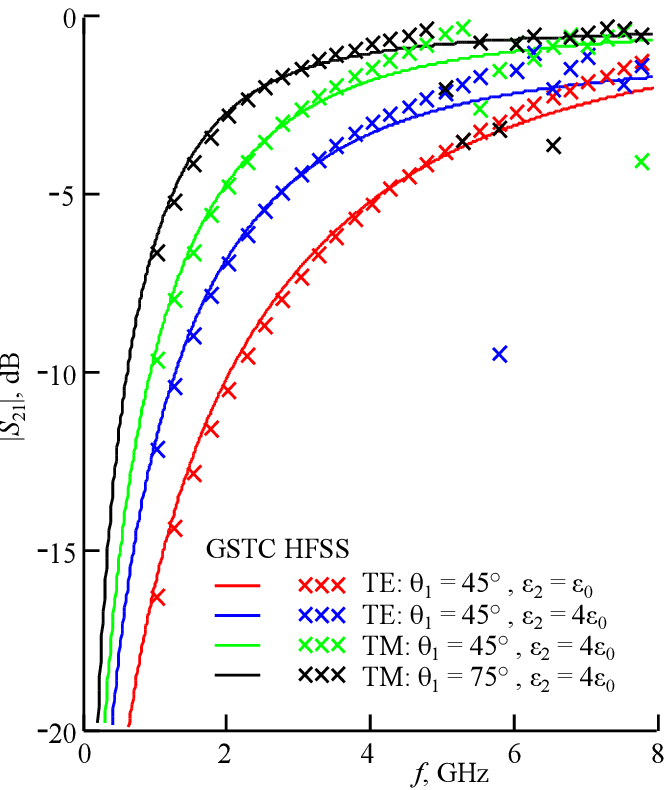}} \\
 (a) \\
 \vspace*{0.1in}
 \scalebox{1.2}{\includegraphics*{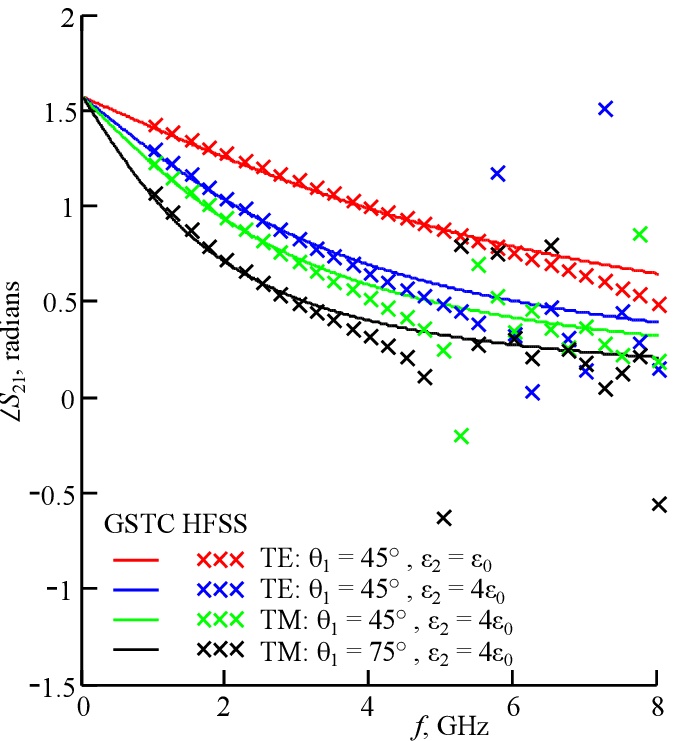}} \\
 (b)
\caption{Oblique incidence transmission coefficients from the metascreen of Fig.~\ref{fsquare} with $a = 18$ mm, $d=20$ mm, $\epsilon_1 = \epsilon_0$ and $\mu_1 = \mu_2 = \mu_0$: (a) magnitude, (b) phase.}
\label{fobl}
\end{figure}

For oblique incidence, Floquet ports were used in the HFSS simulations. In Fig.~\ref{fobl}, we show comparisons between the transmission coefficients obtained numerically and those based on the GSTCs for the metascreen of Fig.~\ref{fsquare}, with various polarizations, angles of incidence and material constants of the second medium.
Results are displayed as $S_{21}$, where
\begin{equation}
 S_{21} = T_{\rm (TE or TM)} \sqrt{\frac{\zeta_1 \cos \theta_2}{\zeta_2 \cos \theta_1}}
\end{equation}
At oblique incidence, the onset of the next higher-order Floquet mode (also known as the grating frequency or Rayleigh frequency \cite{ray}) occurs at
\[
 f_R = \frac{c}{d \left( \sqrt{\epsilon_r} + \sin \theta_1 \right)}
\]
where $\epsilon_r= \epsilon_2/\epsilon_1$. For $\theta_1 = 45^{\circ}$ and $\epsilon_1 = \epsilon_2 = \epsilon_0$, we have $f_{R} = 8.79$ GHz; for $\theta_1 = 45^{\circ}$ and $4\epsilon_1 = \epsilon_2 = 4 \epsilon_0$, we have $f_{R} = 5.54$ GHz; and for $\theta_1 = 75^{\circ}$ and $4\epsilon_1 = \epsilon_2 = 4 \epsilon_0$, we have $f_{R} = 5.06$ GHz. It is observed that for those cases the agreement is very good below $f_{R}/2$, but as $f_{R}$ is approached, significant discrepancies arise (the phase especially deviates wildly above about $0.75 f_{R}$). Note that in the oblique incidence simulations, the finite-element solution mesh had to be much smaller, typically around 20,000 or 30,000 first-order elements, and larger variations in computed $S$-parameters (on the order of a few tenths of a percent) could often be observed at the final mesh refinement, so accuracy better than this for the finite-element simulations should not be assumed.

\section{Conclusion}

In this paper, we have derived GSTCs for a metascreen consisting of a square array of arbitrarily shaped apertures in a perfectly conducting planar screen of vanishing thickness. From these, equivalent shunt reactances have also been obtained. These conditions contain as special cases many such results previously presented in the literature. The boundary conditions contain parameters known as surface porosities that depend only on the shape and spacing of the apertures, and we have presented some formulas for these porosities for several metascreen geometries. The GSTCs enable us easily to derive formulas for reflection and transmission coefficients of plane waves at a metascreen, and their accuracy has been demonstrated by comparison with results of full-wave simulations. We claim that the GSTC model derived in this paper (and extended to metascreens of nonzero thickness in a separate paper \cite{ekel}) holds for apertures of arbitrary shape, provided the general conditions set forth in this paper are satisfied. Validation has been carried out for an array of asymmetric apertures in \cite{hkh}, where retrieval methods for the surface parameters of the GSTCs are also derived..

This work could readily be extended to other array geometries such as rectangular or hexagonal, and it seems likely that frequency-dependent surface porosities could be found that would provide more accuracy when the lattice constant becomes comparable to or larger than half a wavelength, as seen for example in \cite{lee}, \cite{zar} and \cite{medina}. These are subjects for future research.

\appendices

\section{Definition of Aperture Polarizabilities}
\label{apa}
Since various authors have defined the polarizabilities of an aperture in different ways, we present here the definition to be used in this paper, modified slightly from the presentation in \cite{vanbladel}. Let a PEC screen of zero thickness lie in the plane $z=0$, a medium with material parameters $\epsilon_1$, $\mu_1$ occupying the region $z>0$ and one with $\epsilon_2$, $\mu_2$ occupying the region $z<0$. With no aperture in the screen, a set of sources located on both sides of the screen is said to produce the \emph{short-circuit} fields $\mathbf{E}^{\rm sc}$, $\mathbf{H}^{\rm sc}$ (and the corresponding $\mathbf{D}^{\rm sc}$, $\mathbf{B}^{\rm sc}$). When an electrically small aperture is cut into the screen, the total field is equal to the short-circuit field plus an additional field $\mathbf{E}^{\rm ap}$, $\mathbf{H}^{\rm ap}$ due to the aperture. Sufficiently far from the aperture, this field is equal to that produced by a set of electric and magnetic dipoles, \emph{acting in the presence of the 
screen, with the 
aperture closed off (metalized)}. The dipoles acting in the upper half space $z>0$ are located at the center of the aperture at $z=0^+$ and have the values
\begin{equation}
 \mathbf{p}_+ = - \mathbf{u}_z \frac{2\epsilon_1}{\epsilon_1 + \epsilon_2} \alpha_E \left[ D_z^{\rm sc} \right]_{z=0^-}^{0^+} ,
 \label{apa1}
\end{equation}
\begin{equation}
 \mathbf{m}_+ = \frac{2 \mu_2}{\mu_1 + \mu_2} \dyadic{\boldsymbol{\alpha}}_M \cdot \left[ \mathbf{H}^{\rm sc} \right]_{z=0^-}^{0^+}
 \label{apa2}
\end{equation}
while the dipoles acting in the lower half space $z<0$ are located at the center of the aperture at $z=0^-$ and have the values
\begin{equation}
 \mathbf{p}_- = \mathbf{u}_z \frac{2\epsilon_2}{\epsilon_1 + \epsilon_2} \alpha_E \left[ D_z^{\rm sc} \right]_{z=0^-}^{0^+} ,
 \label{apa3}
\end{equation}
\begin{equation}
 \mathbf{m}_- = - \frac{2 \mu_1}{\mu_1 + \mu_2} \dyadic{\boldsymbol{\alpha}}_M \cdot \left[ \mathbf{H}^{\rm sc} \right]_{z=0^-}^{0^+}
 \label{apa4}
\end{equation}
These equations define the polarizabilities $\alpha_E$ and $\dyadic{\boldsymbol{\alpha}}_M$; the magnetic polarizability dyadic has only components in the tangential ($x$ and $y$) directions. Note that the polarizability definitions used here have excluded factors dependent on the material properties on either side of the screen, in contrast, for example, with the definition used in \cite{marin}.

\section{Analytical Expressions for the Surface Porosities}
\label{apb}

All metascreens considered in this appendix have sufficient symmetry that $\pi_{MS}^{xy} = \pi_{MS}^{yx} = 0$ and that $\pi_{MS}^{xx} = \pi_{MS}^{yy} \equiv \pi_{MS}^t$.

\subsection{Small aperture limit}
When the size of the aperture is significantly smaller than the lattice constant $d$, we may use (\ref{epies}) and (\ref{epims}) to obtain analytical expressions for the surface porosities in certain cases. For a circular aperture of radius $r_0$, it is well known that
\begin{equation}
 \alpha_E = \frac{2}{3} r_0^3 ; \qquad \dyadic{\boldsymbol{\alpha}}_M = \frac{4}{3} r_0^3 \left( \mathbf{u}_x \mathbf{u}_x + \mathbf{u}_y \mathbf{u}_y \right)
\end{equation}
Thus,
\begin{equation}
 \pi_{ES}^{zz} = -\frac{2 r_0^3}{3d^2} \frac{1}{1 - \frac{4 r_0^3}{3Rd^2}}
 \label{ecircsmall}
\end{equation}
\begin{equation}
 \pi_{MS}^t = \frac{4 r_0^3}{3d^2} \frac{1}{1 - \frac{4 r_0^3}{3Rd^2}}
 \label{circsmall}
\end{equation}
valid if $r_0 \ll d$. By duality, equations (\ref{ecircsmall})-(\ref{circsmall}) are contained implicitly in the results of Eggimann and Collin \cite{egg}-\cite{collin}; formula (\ref{ecircsmall}) has been given in \cite{grosser}. If the denominator in (\ref{circsmall}) is eliminated, we retrieve a result obtained in, e.~g., \cite{munush}, \cite{kaden} and \cite{delogne}, for which interaction between neighboring apertures is neglected.

For square apertures of side $a$, Fabrikant \cite{fab1}-\cite{fab2} has given the accurate analytical approximations
\begin{equation}
 \alpha_E = \frac{1}{6\sqrt{2}} a^3 ; \qquad \dyadic{\boldsymbol{\alpha}}_M = \frac{2}{9 \ln (1 + \sqrt{2})} a^3 \left( \mathbf{u}_x \mathbf{u}_x + \mathbf{u}_y \mathbf{u}_y \right)
\end{equation}
and therefore, the metascreen shown in Fig.~\ref{fsquare} has the surface porosities
\begin{equation}
 \pi_{ES}^{zz} = -\frac{a^3}{6\sqrt{2}d^2} \frac{1}{1 - \frac{a^3}{3\sqrt{2}Rd^2}}
 \label{apb5}
\end{equation}
\begin{equation}
 \pi_{MS}^t = \frac{2a^3}{9\ln (1 + \sqrt{2})d^2} \frac{1}{1 - \frac{2a^3}{9\ln (1 + \sqrt{2}) Rd^2}}
 \label{apb6}
\end{equation}
valid for $a \ll d$. Fabrikant has also given expressions for the polarizabilities of a number of other shapes, from which further expressions for surface porosities may be obtained (e.~g., for an array of cross-shaped apertures \cite{moller}).

\subsection{Larger apertures}
Equations (\ref{epies}) and (\ref{epims}) can only be expected to be valid when multipole interactions of order higher than dipoles can be neglected---in other words, when  $d$ is significantly larger than the dimensions of the apertures. There are some results in the literature that apply when this condition does not hold. For square apertures whose side length is nearly equal to the lattice constant ($w = d-a \ll d$), the results of Sakurai \cite{sak}, and of Kontorovich and his colleagues \cite{kont2}-\cite{kont4}, give:
\begin{equation}
 \pi_{ES}^{zz} = -\frac{d}{4\pi} \ln \frac{2d}{\pi w}
 \label{apb3}
\end{equation}
\begin{equation}
 \pi_{MS}^t = \frac{d}{2\pi} \ln \frac{2d}{\pi w}
 \label{apb4}
\end{equation}
Related results for the surface electric porosity were obtained earlier in \cite{knoll}-\cite{oertel} in connection with studies of grids in vacuum tubes.

For an array of circular apertures whose diameter can be a larger fraction of the lattice constant, a different formula than (\ref{circsmall}) for the magnetic porosity can be inferred from a comparison of the formula for normal-incidence transmission coefficient given in \cite[equation (11)]{chen} to our formulas (\ref{TTE}) and (\ref{xte}), in which we put $k_1 = k_2 = k_0$ and $\zeta_1 = \zeta_2 = \zeta_0$. In the limit when $k_0 d \ll 1$, this formula becomes
\begin{equation}
 \pi_{MS}^t = \frac{d}{\displaystyle 32 \pi \left\{ \left[ \frac{J_1'(\frac{2\pi r_0}{d})}{1 - \left( \frac{2\pi r_0}{j_{11}' d} \right)^2} \right]^2 + \sqrt{2} \left[ \frac{J_1'(\frac{2\pi r_0}{d} \sqrt{2})}{1 - 2\left( \frac{2\pi r_0}{j_{11}' d} \right)^2} \right]^2\right\}}
 \label{chenn}
\end{equation}
where $J_1$ is the Bessel function of order 1 and $j_{11}' = 1.841 \ldots$ is the smallest positive root of $J_1'$. The formula in \cite{chen} is asserted to be accurate only for $0.28d < r_0 < 0.5d$, and indeed it does not give $\pi_{MS}^t \rightarrow 0$ as $r_0 \rightarrow 0$ as would be expected on physical grounds. Several semi-empirical formulas for an equivalent sheet impedance have been derived by Ramaccia \emph{et al.} \cite[sect. 4.4]{barbuto} (see also \cite{rama1}-\cite{rama2}) for the case when the diameter $2 r_0$ is nearly equal to the lattice constant ($w = d-2r_0 \ll d$). By comparing the resulting formulas for the reflection coefficient of a normally incident plane wave, we can infer from \cite{rama1} the following expression for the magnetic surface porosity:
\begin{equation}
 \pi_{MS}^t = \frac{d}{2\pi} \ln \sec \frac{\pi^2 r_0}{4d}
 \label{rama}
\end{equation}

\begin{figure}[ht]
 \centering
 \scalebox{0.66}{\includegraphics*{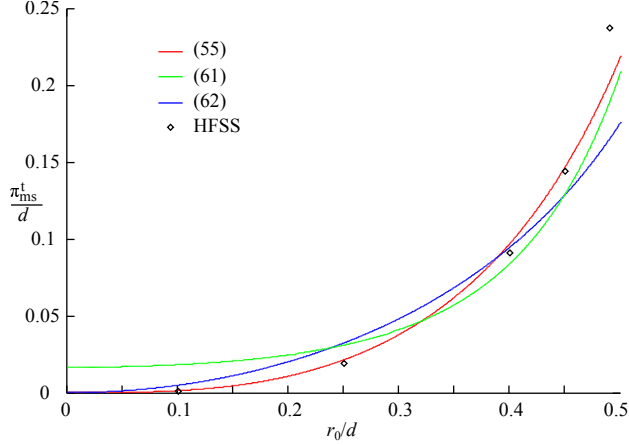}}
\caption{Comparison of three formulas for the magnetic surface porosity of a square array of circular holes.}
\label{f17}
\end{figure}

Values for $\pi_{MS}^t$ can be inferred from full-wave simulation results for the normal incidence transmission coefficient in the case when $\epsilon_1 = \epsilon_2 = \epsilon_0$ and $\mu_1 = \mu_2 = \mu_0$, using (\ref{xte}) and (\ref{TTE}) to obtain
\begin{equation}
 \pi_{MS}^t = \frac{T_{\rm TE}}{2jk_0 (1 - T_{\rm TE})} \qquad \mbox{\rm as $\omega \rightarrow 0$}
 \label{pifromT}
\end{equation}
A comparison of formulas (\ref{circsmall}), (\ref{chenn}) and (\ref{rama}) with results from (\ref{pifromT}) is given in Figure~\ref{f17}, and shows, perhaps surprisingly, that (\ref{circsmall}) is quite accurate over the entire range $0 \leq r_0/d \leq 0.45$, despite being based on a small-hole approximation.
In fact, we see that (\ref{circsmall}) is the best of the three expressions, losing accuracy only when $r_0 \rightarrow 0.5d$, in which case we should expect a percolation threshold $\pi_{MS}^t \rightarrow \infty$ on physical grounds as the boundary condition changes from the form (\ref{egstc}) to that for a metafilm (the first of equations~(1) in \cite{hk3}). No such behavior is exhibited by any of the three formulas. We have not found any closed-form analytical approximations for $\pi_{ES}^{zz}$ in the literature other than (\ref{ecircsmall}).

\subsection{Uniform expressions for square lattice of square apertures}
Following an idea of Grosser and Schulz \cite{grosser} (see also \cite{hlksn}, eqn.~(60)), we can construct interpolated formulas for the metascreen of Fig.~\ref{fsquare} that are uniformly valid for any value of $a/d$. We propose expressions of the form
\begin{equation}
 \pi_{ES}^{zz} = -f_1 \left( \frac{a}{d} \right) \frac{d}{4\pi} \ln \sec \frac{\pi a}{2d}
 \label{apb1}
\end{equation}
\begin{equation}
 \pi_{MS}^t = f_2 \left( \frac{a}{d} \right) \frac{d}{2\pi} \ln \sec \frac{\pi a}{2d}
 \label{apb2}
\end{equation}
where $f_1$ and $f_2$ are functions to be determined.\footnote{
Expression (\ref{apb2}) with $f_2 \equiv 1$ has been given in \cite{matsu}, and is the low-frequency limit of the result obtained in \cite{belyaev}.}
If $f_{1,2} \rightarrow 1$ as $a/d \rightarrow 1$, then (\ref{apb1}) and (\ref{apb2}) will approach the limits given in (\ref{apb3})-(\ref{apb4}) above, because
\begin{equation}
 \sec \frac{\pi a}{2d} = \frac{1}{\sin \frac{\pi w}{2d}} \simeq \frac{2d}{\pi w}
\end{equation}
if $w/d \ll 1$. We now want to choose $f_{1,2}$ in forms as simple as possible so that the limits in (\ref{apb5}) and (\ref{apb6}) are achieved when $a/d \ll 1$. Put
\begin{equation}
 f_1(x) = C_1 x + (1-C_1)x^2
\end{equation}
where $x = a/d$ and $C_1$ is a constant to be determined. This function obeys $f_1(1) = 1$ as required, while $f_1(x) \simeq C_1x$ for $x \ll 1$. Since
\begin{equation}
 \ln \sec \frac{\pi x}{2} \simeq \frac{\pi^2 x^2}{8} \qquad \mbox{\rm for $x \ll 1$}
\end{equation}
we have that
\begin{equation}
 \frac{\pi_{ES}^{zz}}{d} \simeq -\frac{\pi C_1}{32} x^3 \qquad \mbox{\rm for $x \ll 1$}
\end{equation}
and equating this to $x^3/6\sqrt{2}$ from (\ref{apb5}), we get
\begin{equation}
 C_1 = \frac{8\sqrt{2}}{3\pi} = 1.2004 \ldots
\end{equation}
and thus
\begin{equation}
 \frac{\pi_{ES}^{zz}}{d} = -\frac{\ln \sec \frac{\pi a}{2d}}{4\pi} \left[ C_1 \frac{a}{d} + \left( 1 - C_1 \right) \frac{a^2}{d^2} \right]
 \label{apc2}
\end{equation}
is an approximation for $\pi_{ES}^{zz}$ valid uniformly for $0 < a < d$. In a similar manner,
\begin{equation}
 \frac{\pi_{MS}^t}{d} = \frac{\ln \sec \frac{\pi a}{2d}}{2\pi} \left[ C_2 \frac{a}{d} + \left( 1 - C_2 \right) \frac{a^2}{d^2} + \frac{\sin \left( \pi \frac{a^2}{d^2} \right)}{25} \right]
 \label{apc1}
\end{equation}
is a uniform approximation for the magnetic porosity, where
\begin{equation}
 C_2 = \frac{32}{9\pi \ln (1 + \sqrt{2})} = 1.2841 \ldots
\end{equation}
We included the extra term $\sin \left( \pi \frac{a^2}{d^2} \right)/25$ in order to improve the accuracy of (\ref{apc1}), which was verified against full-wave numerical simulation results using (\ref{pifromT}). Both (\ref{apc2}) and (\ref{apc1}) have been compared with results of full-wave numerical simulations (Figs.~\ref{fnorm} and \ref{fobl} are examples of this) and shown to be accurate to within a few percent for any value of $a/d$ between 0 and 1. Note that by complementarity, the result of \cite{hlksn} corresponds to the choice $f_2(x) = \sin \left( \pi x /2 \right)$ in (\ref{apb2}), which was found to give discrepancies with full-wave simulation results about 3 times larger than does (\ref{apc1}), so the latter is to be preferred.

\section*{Acknowledgment} 

The authors are grateful to Dr. Chris Holloway of the National Institute of Standards and Technology for a number of useful discussions about this paper.

\end{document}